\documentclass[prd, aps, nofootinbib, preprintnumbers,
 showpacs, superscriptaddress, twocolumn]{revtex4}

\usepackage{graphicx,epsfig}

\usepackage{amssymb,amsmath}
\usepackage[english]{babel}
\usepackage{graphicx}
\usepackage[latin9]{inputenc}
\usepackage{epstopdf}
\usepackage{color}

\begin{document}

\title{High accuracy simulations of Kerr tails: coordinate dependence and
  higher multipoles}

\author{Manuel Tiglio}

\affiliation{Department of Physics and Astronomy, and Center for Computation and Technology, Louisiana State
  University, Baton Rouge, LA 70803-4001, USA}

\affiliation{Department of Physics, and Center for Scientific Computation and
  Mathematical Modeling, University of Maryland, College Park, MD 20742, USA}

\author{Lawrence E. Kidder}

\affiliation{Center for Radiophysics and Space Research, Cornell University,
  Ithaca, NY 14853, USA.}

\author{Saul A. Teukolsky}

\affiliation{Center for Radiophysics and Space Research, Cornell University,
  Ithaca, NY 14853, USA.}

\begin{abstract}

We investigate the late time behavior of a scalar field on a fixed
Kerr background using a $2+1$ dimensional pseudospectral evolution
code.  We compare evolutions of pure axisymmetric multipoles in both
Kerr-Schild and Boyer-Lindquist coordinates.  We find that the
late-time power-law decay rate depends upon the slicing of the
background, confirming previous theoretical predictions for those decay rates.
The accuracy of the numerical evolutions is sufficient to decide
unambiguously between competing claims in the literature.

\end{abstract}


\maketitle

\section{Introduction}

Following the original work of Price \cite{Price1972}, 
the late time behavior of fields propagating on a Schwarzschild black
hole has been studied for decades. By now there is a
rather complete theoretical picture of it, with compelling supporting
numerical evidence. Since in the Schwarzschild case the background
geometry is spherically symmetric, perturbations can be expanded in
spherical harmonics. Solutions with different multipole numbers
$(\ell,m)$ evolve independently. For initial data that is not time
symmetric, each of these perturbations decays at late times as
$t^{-(2\ell +3)}$, regardless of the type of perturbation (scalar,
gravitational, electromagnetic, etc).

In contrast, our understanding of the late-time decay on a Kerr
background is still incomplete. In fact, it has been the subject of
controversy over the past decade.

Since in the spinning case the background is not spherically
symmetric, additional multipoles are generated during the evolution of
an initially pure multipole. The first analytical calculations for the
decay rate of these dynamically generated multipoles were done by
Barack and Ori \cite{Barack1999} and Hod \cite{Hod2000}.
According to their calculations,
the interactions between different multipoles and the
black hole angular momentum change the decay rate of each of those
multipoles compared to the Schwarzschild case.  Furthermore, those
calculations also predict that in the Kerr case the late time decay
rate is not universal, in the sense that it depends on the type of
perturbations.

For example, Hod \cite{Hod2000}
predicted that a scalar perturbation with an initial
pure multipole structure with indices $(\ell,m)$ on a Kerr background
in Boyer-Lindquist coordinates would be dominated at late times by the
following multipole component and decay:
\begin{eqnarray}
\Psi & \propto & Y^{(\ell,m)}t^{-(2\ell+3)} \quad \mbox{ if } \ell=m \mbox{ or }
\ell=m+1 , \\
\Psi & \propto & Y^{(\ell=m,m)}t^{-(\ell+m+1)} \quad \mbox{ if } \ell-m \ge 2
\mbox{ (even)} , \\
\Psi & \propto & Y^{(\ell=m+1,m)}t^{-(\ell+m+2)} \quad \mbox{ if } \ell-m \ge 2
\mbox{ (odd)}.
\end{eqnarray}

In contrast, Burko and Khanna \cite{Burko2002} argued in
favor of a ``simple picture'', in which at late times the decay would
also be dominated by the lowest multipole that can be generated by
mode mixing during the
evolution, but that such a mode decays in the same way that it would on
a Schwarzschild background. In other words, the authors claimed that
the details of the multipole interactions do not affect the late time
decay rate. Therefore, according to this simple picture the late time
decay for a massless scalar field in a spinning Kerr background should
simply be $t^{-(2\ell_{\rm min} +3)}$, where $\ell _{\rm min}$ is the lowest
multipole that can be generated during the evolution. For example,
in the axisymmetric case an initial perturbation that is symmetric
about the equator would produce even $\ell$ modes by mixing, with $\ell_{\rm min}
=0$. An antisymmetric initial perturbation would have $\ell_{\rm min}=1$.

The simplest case in which these predictions differ is the
axisymmetric solution of the wave equation with $\ell=4$ initial
data. The lowest multipole that can be generated during evolution is
the monopole term. The simple picture argument predicts for it a decay
as in Schwarzschild, $t^{-3}$, while Hod's calculations predict
instead a decay of $t^{-5}$. Early numerical studies of the wave
equation on a Kerr background in Boyer-Lindquist coordinates by Krivan
\cite{Krivan99} found an approximate decay of $t^{-5.5}$, which was
interpreted by some as an approximate verification of Hod's
prediction.  Others suggested that Krivan's results were plagued by
numerical problems and that Krivan's time decay of $t^{-5.5}$ should
not be interpreted as supporting Hod's prediction,
but as a numerical artifact, a transient, or both
\cite{Burko2002}. Krivan in fact, had reported problems with the
angular differentiation in the code used in \cite{Krivan99}. In particular 
the decay rate found in \cite{Krivan99} does not
appear to get closer to $t^{-5}$ with increasing angular resolution
but to $t^{-5.5}$, leaving room for questioning the interpretation of
the result as a confirmation of Hod's prediction.

In order to settle this issue, Burko and
Khanna \cite{Burko2002}
performed simulations of the wave equation on a Kerr background
using ``ingoing Kerr'' coordinates and obtained a late time decay
very close to $t^{-3}$, concluding that Krivan's results were indeed
misleading because of numerical artifacts.

It turns out, however, that the simulations of Burko and Khanna and
Krivan do not correspond to the same physical scenario. In the
spinning case, the characterization of the initial data in terms of a
spherical harmonic decomposition is not unique, since there is no
preferred sphere with respect to which such a decomposition should be
done. In particular, a pure $\ell=4$ initial perturbation in
ingoing Kerr coordinates does not correspond to pure $\ell=4$
initial data in Boyer-Lindquist coordinates, and vice-versa. Not only
are the spatial coordinates different, but also the slicing of
spacetime. In fact, in a more recent analytic calculation, Poisson
\cite{Poisson2002}
explicitly showed that for perturbations of flat spacetime (where, in
particular, the contribution due to the spin is treated only to leading
order) the decay does depend on the choice of coordinates.
Furthermore, Poisson was able to obtain both Hod's
and the simple prediction within this linearized scenario, depending
on which coordinates are used.

In Ref.\ \cite{Scheel2004} Scheel et al.\ presented high accuracy
evolutions of the wave equation on a Kerr background using pure
multipole initial data in Kerr-Schild coordinates up to $\ell=4$.
Their numerical results confirmed the predictions of the simple
picture with high accuracy; in particular, they were consistent with
those of Burko and Khanna.  

More recently, Gleiser, Price and Pullin \cite{Gleiser2007}
have presented numerical
results for linear scalar perturbations of a Schwarzschild black hole
where the effect of the spin is treated perturbatively. This is done
through a hierarchy of one-dimensional evolution equations, which take
into account increasingly higher order corrections due to the spin, in
what would correspond to Boyer-Lindquist coordinates. Their results
show that if one starts with a pure $\ell=4$ mode in those coordinates
and solves the corresponding hierarchy of equations to study the late
time behavior of the monopole term (which is the one that dominates
at late times) at the leading order in the angular momentum, it
decays as in Hod's prediction (i.e., as $t^{-5}$), not as in the
simple picture prediction. The advantage of this approach is that the
resulting equations can be solved
by a simple numerical code, since they are one-dimensional in space.
The disadvantage is that the treatment is
perturbative in the black hole angular momentum, leaving open the
question of whether the full non-linear dependence on the angular
momentum would change the results.

The goal of this paper is to study the late time behavior of fields in
a Kerr spacetime by numerically evolving a scalar field in
axisymmetry, including the full dependence of the equations on the
spin, and unambiguously determine what the decay is, and whether it
depends on the coordinates used. For that purpose we numerically
evolve a scalar field on a Kerr background, using both Kerr-Schild and
Boyer-Lindquist coordinates. Our simulations are of enough
accuracy so as to rule out any possibility of
numerical artifacts. We discuss the case that has been under dispute,
which is the simplest one in which Hod's and the simple prediction
disagree, namely pure $\ell=4$ initial data. We also go beyond it
and analyze the $\ell=5,6,7,8$ initial data cases.

Our results for the Kerr-Schild case extend those of Scheel et
al.\ and Burko and Khanna to higher accuracy and higher multipole
initial data. In particular, we confirm that the decay in those
coordinates is governed by the predictions of the ``simple
picture''. For Boyer-Lindquist coordinates our results extend those of
Krivan and of Gleiser, Pullin and Price, again to higher accuracy and
higher multipole initial data. Consistent with the conclusions of
those references, we unambiguously find that the decay rate in these
coordinates is the one predicted by Hod.

The organization of this paper is as follows.  In
Sec.~\ref{sec:numerical-method} we describe the method we use to
numerically evolve the scalar field.  In Sec.~\ref{sec:results} we
present our results for pure $\ell$ initial data for both Kerr-Schild
and Boyer-Lindquist coordinates.  Finally, we present our conclusions
in Sec.~\ref{sec:conclusions}.

\section{The numerical method}
\label{sec:numerical-method}

We evolve axisymmetric
solutions of the wave equation $\nabla ^a \nabla _a \Psi = 0$ on a
fixed Kerr black hole background in both 
Kerr-Schild and Boyer-Lindquist coordinates. In both cases we cast and
numerically solve the equations as a first order in space
and time system. For example, in the Boyer-Lindquist case we write the
equations as
\begin{eqnarray*}
\dot{\Psi}_t &=& \frac{(r^2+a^2)^2}{\Sigma^2} \partial_{r_*} \Psi _{r_*} +
 \frac{2r\Delta}{\Sigma ^2} \Psi _{r_*}  - \frac{2y\Delta}{\Sigma ^2}\Psi _y \\
&& +\frac{\Delta(1-y^2)}{\Sigma ^2} \partial_y \Psi _y , \\
\dot{\Psi}_{r_*} &=& \partial_{r_*} \Psi _t , \\
\dot{\Psi}_y &=& \partial_y \Psi _t , \\
\dot{\Psi} &=& \Psi_t ,
\end{eqnarray*}
where $M,a$ are the mass and spin of the black hole, respectively, 
\begin{eqnarray*}
\Delta &=& r^2-2Mr+a^2 , \\
\Sigma &=& \left[ (r^2+a^2)^2 - a^2\Delta(1-y^2) \right]^{1/2} , \\
y &=& -\cos {\theta} ,
\end{eqnarray*}
 and where we have introduced the Kerr-tortoise coordinate $r_*$, defined
by
$$
\frac{dr_*}{dr} = \frac{r^2+a^2}{\Delta} \, .
$$
Here $\Psi _{r_{*}}$, $\Psi _t$ , and $\Psi _y$ are auxiliary variables
 introduced to cast the system in first order form; at the
continuum they satisfy $\Psi _{r_{*}}= \partial _{r_*} \Psi$, $\Psi _t
=\partial_t \Psi$, and $\Psi _y \ = \partial _y \Psi$.

We solve the resulting equations using a special purpose two-dimensional code written for
this project. A requirement for the numerical solution is that it be
demonstrably accurate enough that there is no doubt about the results.
Since the solution is smooth, a spectral method should be optimal in
terms of efficiency for high accuracy. Accordingly, we use a pseudo-spectral
collocation (PSC) method
in space and the method of lines to evolve in time.
A by-product of using a spectral method is that we avoid all difficulties
in handling the polar singularities in spherical coordinates.

In the simulations shown below the domain in the radial direction is
partitioned into blocks, each of length $10M$ (in either
Boyer-Lindquist-tortoise or Kerr-Schild coordinates). On each block the
dependence of the solution in the angular and radial directions is
expanded in spherical harmonic and Chebyshev polynomials,
respectively, using Gauss-Lobatto collocation points. The solution is
advanced in time at each of these points using a fourth-order
Runge-Kutta method. Information at the interfaces of the different blocks is
communicated through
characteristic variables using a penalty method, as described in
\cite{Hesthaven1997,Hesthaven1999}. 

We use initial data of the form
\begin{eqnarray*}
\Psi(t=0) &=& 0 , \\
\Psi_t(t=0) &=& e^{-\left( r-r_0 \right) ^2/\sigma ^2 } Y^{(\ell,0) }, 
\end{eqnarray*}
where $r_0=20M$ and $\sigma = 4M$.
In these expressions as well as in the results shown in the next section, 
 $r$ and $t$ are the
Boyer-Lindquist or Kerr-Schild radial and time coordinates, respectively, depending on the
equations being solved.

We set the angular momentum to $a=0.5M$. In the Kerr-Schild case the
black hole singularity is excised by placing a purely outflow inner
boundary at $1.8M$, while in the Boyer-Lindquist case the inner
boundary is placed at $r_*=-40M$ (which in Boyer-Lindquist radius 
corresponds to a distance of $\sim 10^{-8}M$ from the event
horizon) and we set the incoming modes to zero. In both cases the
outer boundary is placed far away enough so that the results here shown are
causally disconnected from the type of boundary conditions there
imposed (incoming modes set to zero). In more detail, if we evolve for
a total time $T$, we typically place the outer boundary at
$r=T/2+100M + r_{IB}$, where $r_{IB}$ is the radial location of the inner boundary.

The number of collocation points in the radial and angular directions
per domain is denoted by $n_r,n_{\ell}$, respectively and, unless
otherwise stated, the time step is kept fixed at $\Delta t=0.025M$.
Obtaining fast convergence as we increase the number of
collocation points shows that the time step is small enough that the
errors due to the time integration are smaller than those associated
with the spatial dimensions.

In order to be able to follow the solution for long enough periods of
time, especially for the higher multipole initial data cases (which
decay faster) we use quadruple precision.

\section{Results}
\label{sec:results}

Hod's predictions and the simple one coincide for $\ell=0,1,2,3$ initial
data. We have checked that our simulations reproduce the expected
decay for each of those values of $\ell$, using both Boyer-Lindquist
and Kerr-Schild coordinates. In particular, we have found that $\ell=0$
and $\ell=1$ modes {\em already present} in the initial data have a late time decay of $t^{-3}$ and $t^{-5}$,
respectively. As we will show below, and as predicted by Hod and Poisson, the decay of
these multipoles is different when they are {\em dynamically generated} in a
Boyer-Lindquist background.

In the following we report our results
for higher multipole initial data, for which Hod's prediction and the simple
one disagree. We show our results for observers at $r=21.8M$ in the Kerr-Schild
case and $r_*=20M$ in the Boyer-Lindquist one, though similar results hold for
other observer locations.

\subsection{The $\ell=4$ case}

\begin{figure}
 \includegraphics[width=0.48\textwidth]{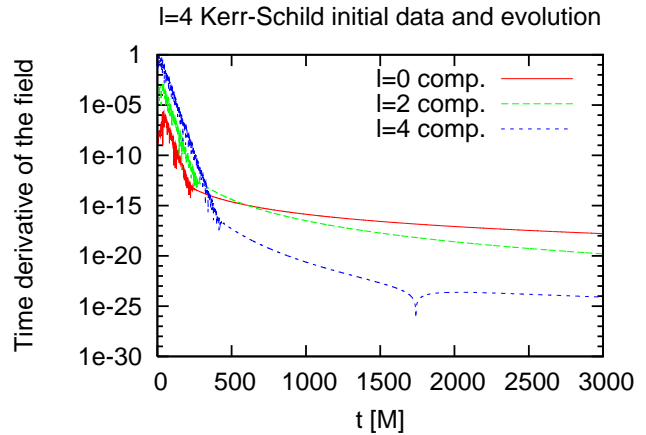}
  \caption{Decay of an initial $\ell=4$ mode in Kerr-Schild coordinates.}
  \label{fig:l4_KS}
\end{figure}

\begin{figure}
  \includegraphics[width=0.48\textwidth]{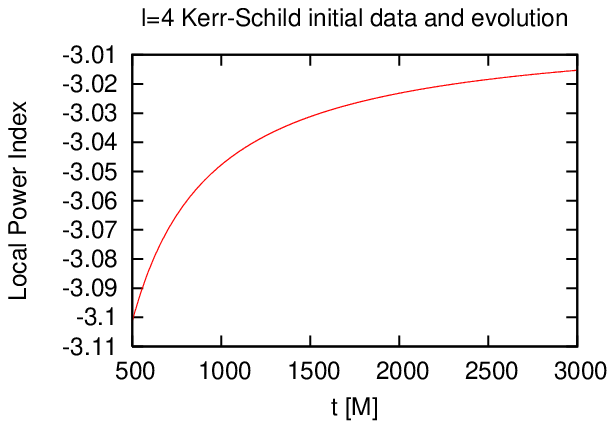}
  \caption{Local power index for the monopole term of the previous figure.}
  \label{fig:l4_KS_LPI}
\end{figure}

Figure \ref{fig:l4_KS} shows the decay of the scalar field versus time for a 
fixed observer radius and $\ell=4$ initial data and evolution in Kerr-Schild
coordinates. Shown are the first few ($\ell=0,2,4$) even multipole components
of the solution [the odd components
stay at quadruple precision roundoff values ($\sim 10^{-35}$) at
all times]. Initially only the $\ell=4$ mode is present, but additional modes
are generated during evolution; at late times
the monopole term dominates. 

Figure \ref{fig:l4_KS_LPI} shows
the local power index, defined as $LPI(\ell,t) = -t\Psi^{(\ell
  )}_t/\Psi ^{(\ell )}$ (where $\Psi ^{(\ell )}$ is the projection of the
solution to its $\ell$-multipole component), 
for the monopole term shown in Fig.\ref{fig:l4_KS}.  If the monopole term 
decays as $\Psi ^{(0)} \propto t^{-\mu }$ at late times, then one should
obtain that $LPI \rightarrow \mu $. The LPI for the monopole is
approaching $-3$ at late times in our simulations, as predicted by the simple picture.

\begin{figure}
  \includegraphics[width=0.48\textwidth]{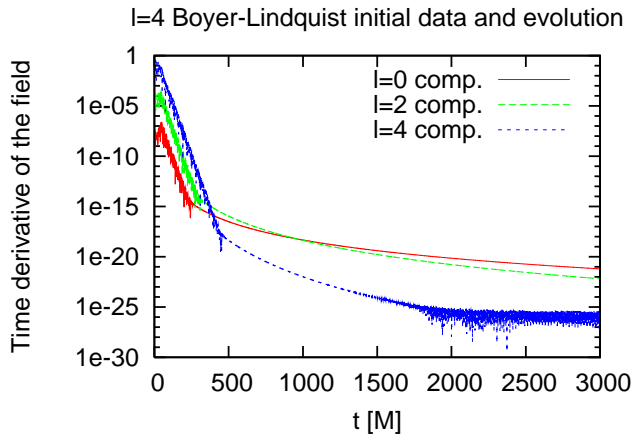}
  \caption{Decay of an initial $\ell=4$ mode in Boyer-Lindquist coordinates.}
  \label{fig:l4_BL}
\end{figure}

\begin{figure}
  \includegraphics[width=0.48\textwidth]{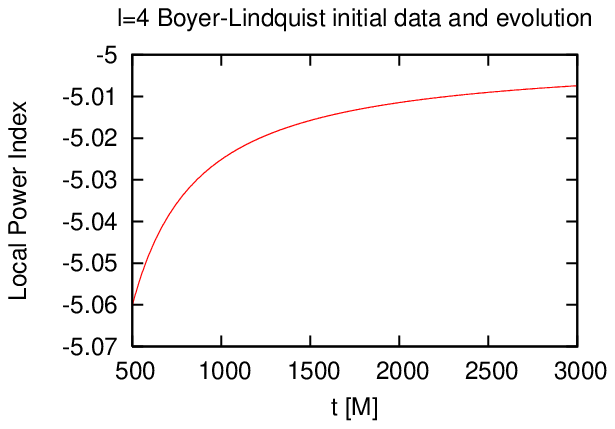}
  \caption{Local power index for the monopole term of the previous
    figure.}
  \label{fig:l4_BL_LPI}
\end{figure}

\begin{figure}
  \includegraphics[width=0.48\textwidth]{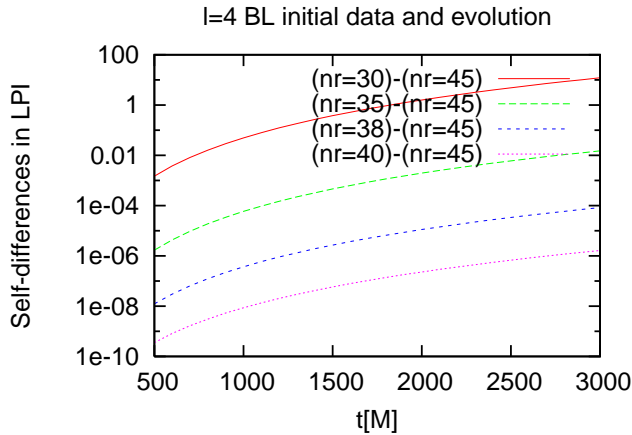}
 \includegraphics[width=0.48\textwidth]{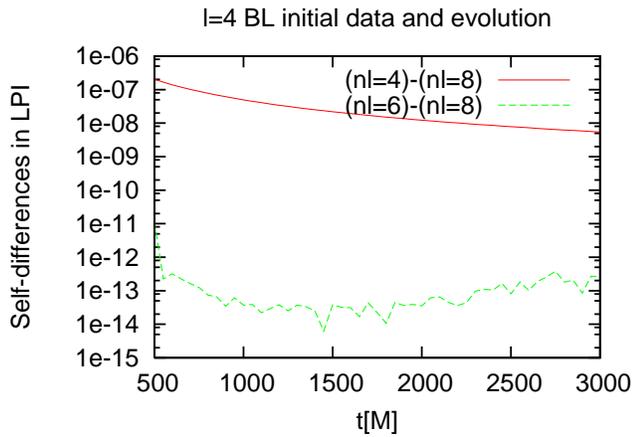}
  \caption{Errors in the LPI for the monopole term of the previous figure. }
  \label{fig:l4_BL_LPI_conv}
\end{figure}

Figure \ref{fig:l4_BL} shows the decay for the evolution of $\ell=4$
initial data in Boyer-Lindquist coordinates. The slowest decaying multipole is
again the monopole term, but in this case we find that it decays as $t^{-5}$, as predicted by Hod and shown in
Fig.~\ref{fig:l4_BL_LPI}. 

The results shown above were obtained with $n_r=45$ and $n_{\ell}=4$.
To give an idea of the errors in these
simulations, Fig.~\ref{fig:l4_BL_LPI_conv} shows the differences in
the LPI for the monopole term between different spatial resolutions ($n_r=30,35,38,40$) and
the highest one ($n_r=45$) as a function of time, keeping $n_{\ell}=4$
fixed.  Similarly, the figure also shows the differences in the LPI
when keeping $n_r=45$ fixed and increasing $n_{\ell}$. At late times
the errors introduced by using $n_{\ell}$ Legendre polynomials, where
$\ell$ is the multipole index of the initial data (as opposed to using
higher order polynomials) is comparable to the errors in the radial
direction. Therefore, unless otherwise specified, in the simulations below we use $n_{\ell}$
Legendre polynomials, where the value of $\ell$ is the same as in the
initial data, and $n_r=45$.

\subsection{The $\ell=5$ case}
\begin{figure}
\includegraphics[width=0.48\textwidth]{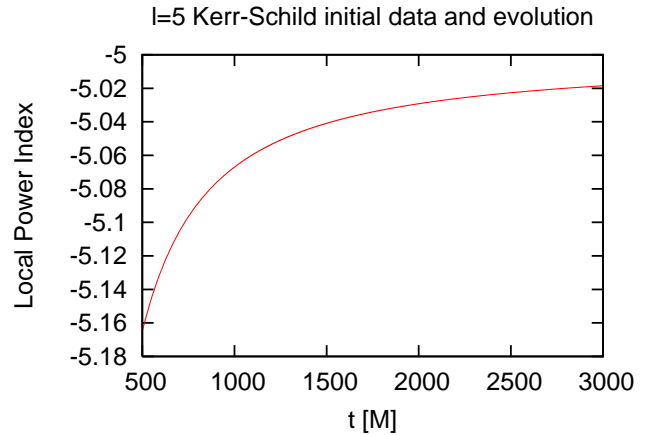}
\includegraphics[width=0.48\textwidth]{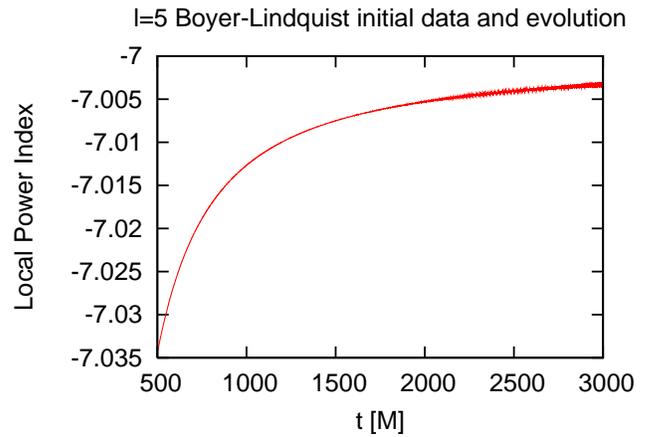}
  \caption{Local power index for the dipole term of $\ell=5$ 
    Kerr-Schild and Boyer-Lindquist initial data and evolution.}
  \label{fig:l5}
\end{figure}

Figure \ref{fig:l5} shows our results for the dipole LPI in evolutions
of $\ell=5$ initial data in Kerr-Schild and Boyer-Lindquist 
coordinates. In both cases the dipole term eventually dominates, with
a decay of $t^{-5}$ in the Kerr-Schild case, as predicted by the
``simple picture'', and a decay of $t^{-7}$ in Boyer-Lindquist
coordinates, as predicted by Hod.

\subsection{The $\ell=6$ case}

\begin{figure}
 \includegraphics[width=0.48\textwidth]{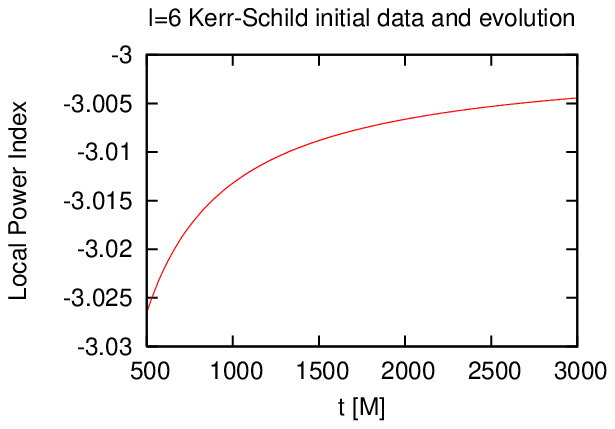}
 \includegraphics[width=0.48\textwidth]{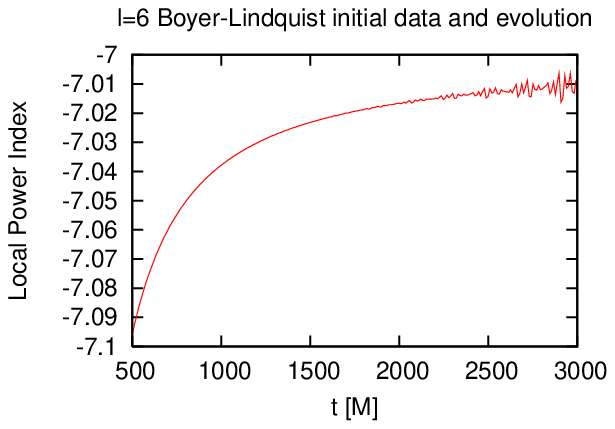}
  \caption{Local power index for the monopole term of $\ell=6$ 
  Kerr-Schild and Boyer-Lindquist initial data and evolution.}
  \label{fig:l6}
\end{figure}

Figure \ref{fig:l6} shows our results for the monopole LPI in
evolutions of $\ell=6$ initial data in Kerr-Schild and Boyer-Lindquist 
coordinates. According to the simple picture interpretation, the
monopole term should dominate at late times, with a decay of $t^{-3}$,
while Hod's prediction for this case is a decay of $t^{-7}$.

\subsection{The $\ell=7$ case}

\begin{figure}
 \includegraphics[width=0.48\textwidth]{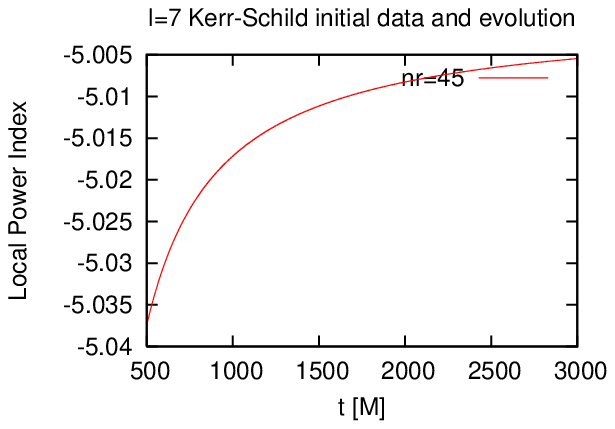}
 \includegraphics[width=0.48\textwidth]{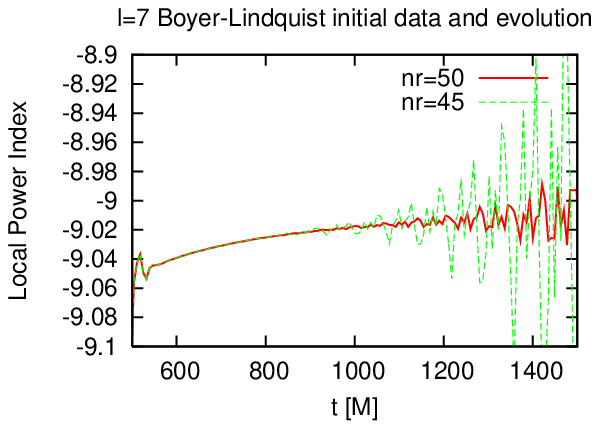}
  \caption{Local power index for the dipole term of $\ell=7$ 
  Kerr-Schild and Boyer-Lindquist initial data and evolution.}
  \label{fig:l7}
\end{figure}

Figure \ref{fig:l7} shows our results for the dipole LPI
in evolutions of $\ell=7$ initial data in Kerr-Schild and Boyer-Lindquist 
coordinates. In the latter case the solution decays much faster and we
need to use higher resolution in the radial direction, with an associated 
smaller timestep for CFL stability. Figure \ref{fig:l7} shows our results for
both $n_r=45,\Delta t=0.025$ and $n_r=50,\Delta t=0.0125$. According
to the simple picture interpretation, the dipole term should
dominate at late times, with a decay of $t^{-5}$, while Hod's
prediction for this case is a decay of $t^{-9}$.

\subsection{The $\ell=8$ case}

\begin{figure}
 \includegraphics[width=0.48\textwidth]{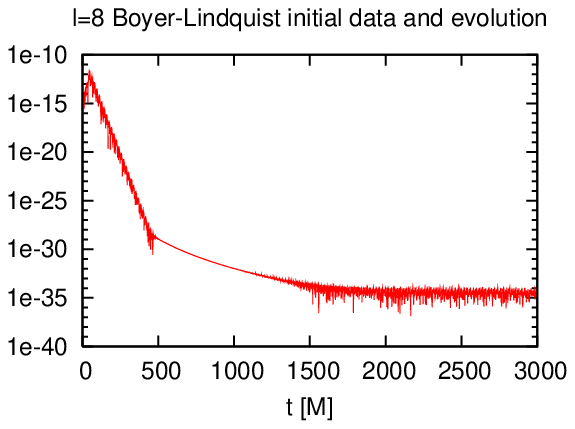}
  \caption{Decay of the monopole term for $\ell=8$ Boyer-Lindquist initial data and evolution.}
  \label{fig:l8_BL}
\end{figure}

\begin{figure}
 \includegraphics[width=0.48\textwidth]{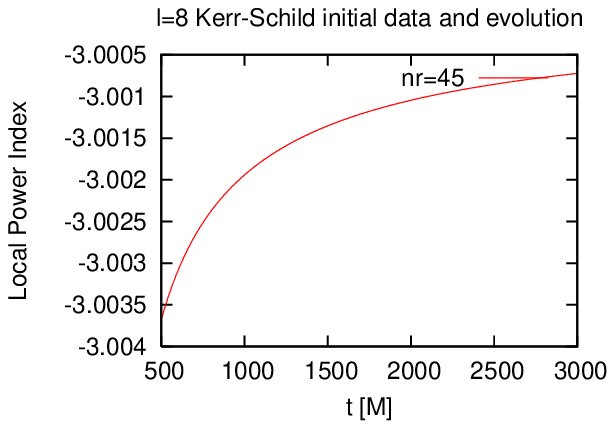}
 \includegraphics[width=0.48\textwidth]{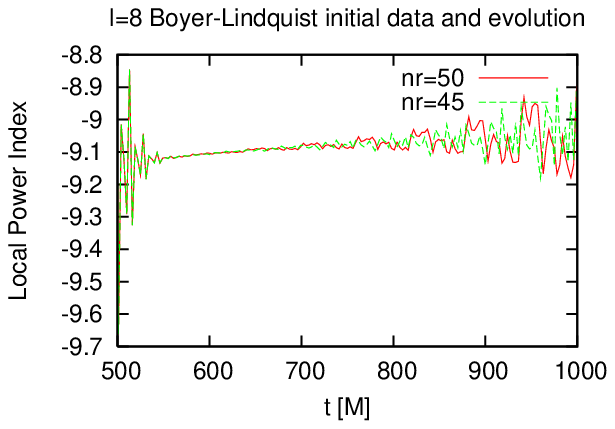}
  \caption{Local power index for the monopole term of $\ell=8$ 
  Kerr-Schild and Boyer-Lindquist initial data and evolution.}
  \label{fig:l8}
\end{figure}

In the Boyer-Lindquist case the monopole term decays even faster for $\ell=8$ initial
data and we start reaching quadruple precision roundoff values by the
time the tail regime begins, as seen in Figure \ref{fig:l8_BL}. As a
consequence, there is a rather short time interval in which we can
measure the local power index, and the latter is not as clean as for
the lower multipoles. Figure \ref{fig:l8} shows our results for
evolutions of $\ell=8$ initial data in both Kerr-Schild and Boyer-Lindquist 
coordinates. According to the simple picture
interpretation, the monopole term should dominate at late times, with
a decay of $t^{-3}$, while Hod's prediction for this case is a decay
of $t^{-9}$.

\section{Discussion}
\label{sec:conclusions}

In this paper we have evolved an axisymmetric scalar field 
on a Kerr background using both Boyer-Lindquist and Kerr-Schild coordinates,
with pure multipole initial data with indices $\ell=0,1,2 \ldots 6,7,8$, and established
the decay rate at late times for each initial data and choice of
background coordinates. 
The numerical values that we obtained for those rates 
confirm the ``simple picture'' prediction \cite{Burko2002} in the Kerr-Schild
case, but Hod's prediction \cite{Hod2000} in
the Boyer-Lindquist one.
Thus, the most important conclusion of this paper is that the observed
late-time decay of the field depends on the time-slicing of the background
spacetime.

The differences between the decay rates that we found in our simulations and the above asymptotic
($t \rightarrow \infty $) predictions are small enough so as to rule out the possibility of numerical
artifacts. Those differences are typically less than one percent, except for the
$\ell =8$ initial data Boyer-Lindquist case. The dominant factor in these differences in
all cases appears to be the fact that we are evolving for a long but finite time; the numerical errors
in our simulations are several order of magnitudes smaller. In the particular
case of Boyer-Lindquist evolutions of $\ell =8$ initial data, the solution reaches quadruple precision roundoff errors
while entering the tail regime. For that reason we cannot determine the
decay rate with uncertainties as small as in the $\ell =0 \ldots 7$ cases. The
late time decay rate that we find for the $\ell=8$ initial data Boyer-Lindquist case 
 is roughly between $-9.2$ and $-9.0$, to be compared with Hod's prediction ($-9$) and
 the simple picture one ($-3$).

Our results are in apparent, but not real, contradiction with those of Scheel
et al. \cite{Scheel2004} and recent work by Burko and Khanna \cite{Burko2007}. In order to clarify the source of
these apparent contradictions we need to describe parts of those references in some
detail.  

The 3D code used by Scheel et al. in Ref.~\cite{Scheel2004} only allowed
for evolutions on a black hole background where the singularity is
excised from the computational domain by placing a purely outflow
inner boundary.  Since this excludes Boyer-Lindquist coordinates, 
Kerr-Schild ones $(t,x,y,z)$ were used in Ref.~\cite{Scheel2004}.  
In this coordinate system, the metric is
given by
\[ g_{\mu \nu} = \eta_{\mu \nu} + 2 H \ell_\mu \ell_\nu, \]
where
\begin{eqnarray*}
H &=& \frac{M r^3}{r^4 + a^2 z^2} , \\
\ell_\mu &=& \{1,\frac{r x + a y}{r^2 + a^2},\frac{r y - a x}{r^2 + a^2},
\frac{z}{r} \},
\end{eqnarray*}
and the radial coordinate $r$ is defined not as a spherical coordinate
but rather a spheroidal coordinate:
\[ \frac{x^2 + y^2}{r^2 + a^2} + \frac{z^2}{r^2} = 1 . \]
Scheel et al. evolved scalar waves on a Kerr-Schild
background using two sets of initial data, corresponding to pure multipoles in
spherical and spheroidal coordinates on constant-time Kerr-Schild slices.  In both cases,
they found the tail decayed as in the simple picture.  Our Kerr-Schild
evolutions confirm this.  
However, this does not necessarily
correspond to the evolution of  pure multipole initial data on a constant-time
Boyer-Lindquist slice, as seen by a Boyer-Lindquist observer. Hod's prediction is for
the late time decay rate seen by an observer with constant
Boyer-Lindquist radius as a function of Boyer-Lindquist time, for an
evolution of pure multipole initial data in Boyer-Lindquist
coordinates. This is exactly what we have explicitly numerically
solved for in this paper.

In more recent work \cite{Burko2007}, Burko and Khanna evolved pure
multipole $\ell =4$ initial data, using Boyer-Lindquist coordinates to
define the multipole decomposition as well as to perform the
evolution, obtaining a decay of $t^{-3}$, which appears to be
consistent with the simple picture and in contradiction with our
results. The reason for this apparent contradiction seems to be caused
by the type of initial data used in Ref.~\cite{Burko2007}. Rather
than giving pure multipole initial data to the scalar field and its
time derivative as done in Hod's original work and in this paper, in
Ref.~\cite{Burko2007} pure multipole initial data was given to the
scalar field and its ``momentum'',
\begin{eqnarray*}
\Psi (t=0) &=& g(r) Y^{(\ell ,m)} ,\\ \left( \partial _t + b \partial
_{r_*} \right )\Psi (t=0) &=& 0,
\end{eqnarray*}
where $b = b(r,\theta)$. The angular dependence of $b$ effectively
corresponds to adding an $\ell =0$ component to the initial data for
$\partial_t \Psi $.  As explained at the beginning of
Sec.~\ref{sec:results}, the decay rate of an initial $\ell = 0$ mode
for the pair $( \Psi, \partial_t \Psi )$ according to both Hod's
prediction and the simple picture is $t^{-3}$ (our simulations, as
well as previous ones, have in particular confirmed this).  What we
have found in this paper, though, which is also reported by Gleiser,
Price and Pullin \cite{Gleiser2007}, is that the decay rate of a {\em
dynamically generated} monopole in a Boyer-Lindquist coordinates is
faster. If as initial data for $( \Psi, \partial_t \Psi )$ one
superposes $\ell =4$ and $\ell =0$ modes, as effectively done in
Ref.~\cite{Burko2007}, the late time decay rate of each component can
be considered independently, since the evolution equations considered
are linear. The evolution of the initial $\ell =4$ component will
dynamically generate a monopole term that decays as $t^{-5}$, while
the evolution of the initial $\ell=0$ component will be dominated by a
decay rate of $t^{-3}$. The results of Ref.~\cite{Burko2007} can be
understood by the fact that the latter will dominate over the former
at late times.

\begin{acknowledgments}

This research was supported in part by NSF grant PHY 0505761 to
Louisiana State University, by a grant from the Sherman Fairchild
Foundation to Cornell, and by NSF grants PHY-0652952, DMS-0553677,
PHY-0652929, and NASA grant NNG05GG51G at Cornell. The research
employed the resources of the CCT at LSU, which is supported by
funding from the Louisiana legislature's Information Technology
Initiative. 

MT would like to thank Enrique Pazos and Jorge Pullin for helpful discussions
throughout this project.

\end{acknowledgments}

\bibliography{References/References}

\begin{thebibliography}{11}
\expandafter\ifx\csname natexlab\endcsname\relax\def\natexlab#1{#1}\fi
\expandafter\ifx\csname bibnamefont\endcsname\relax
  \def\bibnamefont#1{#1}\fi
\expandafter\ifx\csname bibfnamefont\endcsname\relax
  \def\bibfnamefont#1{#1}\fi
\expandafter\ifx\csname citenamefont\endcsname\relax
  \def\citenamefont#1{#1}\fi
\expandafter\ifx\csname url\endcsname\relax
  \def\url#1{\texttt{#1}}\fi
\expandafter\ifx\csname urlprefix\endcsname\relax\def\urlprefix{URL }\fi
\providecommand{\bibinfo}[2]{#2}
\providecommand{\eprint}[2][]{\url{#2}}

\bibitem[{\citenamefont{{Price}}(1972)}]{Price1972}
\bibinfo{author}{\bibfnamefont{R.~H.} \bibnamefont{{Price}}},
  \bibinfo{journal}{\prd} \textbf{\bibinfo{volume}{5}}, \bibinfo{pages}{2419}
  (\bibinfo{year}{1972}).

\bibitem[{\citenamefont{Barack and Ori}(1999)}]{Barack1999}
\bibinfo{author}{\bibfnamefont{L.}~\bibnamefont{Barack}} \bibnamefont{and}
  \bibinfo{author}{\bibfnamefont{A.}~\bibnamefont{Ori}},
  \bibinfo{journal}{Phys. Rev. Lett.} \textbf{\bibinfo{volume}{82}},
  \bibinfo{pages}{4388} (\bibinfo{year}{1999}).

\bibitem[{\citenamefont{Hod}(2000)}]{Hod2000}
\bibinfo{author}{\bibfnamefont{S.}~\bibnamefont{Hod}}, \bibinfo{journal}{Phys.
  Rev.} \textbf{\bibinfo{volume}{D61}}, \bibinfo{pages}{024033}
  (\bibinfo{year}{2000}).

\bibitem[{\citenamefont{Burko and Khanna}(2003)}]{Burko2002}
\bibinfo{author}{\bibfnamefont{L.~M.} \bibnamefont{Burko}} \bibnamefont{and}
  \bibinfo{author}{\bibfnamefont{G.}~\bibnamefont{Khanna}},
  \bibinfo{journal}{Phys. Rev. D} \textbf{\bibinfo{volume}{67}},
  \bibinfo{pages}{081502} (\bibinfo{year}{2003}).

\bibitem[{\citenamefont{Krivan}(1999)}]{Krivan99}
\bibinfo{author}{\bibfnamefont{W.}~\bibnamefont{Krivan}},
  \bibinfo{journal}{Phys.\ Rev.\ D} \textbf{\bibinfo{volume}{60}},
  \bibinfo{pages}{101501} (\bibinfo{year}{1999}).

\bibitem[{\citenamefont{Poisson}(2002)}]{Poisson2002}
\bibinfo{author}{\bibfnamefont{E.}~\bibnamefont{Poisson}},
  \bibinfo{journal}{Phys.\ Rev.\ D} \textbf{\bibinfo{volume}{66}},
  \bibinfo{pages}{044088} (\bibinfo{year}{2002}).

\bibitem[{\citenamefont{Scheel et~al.}(2004)\citenamefont{Scheel, Erickcek,
  Burko, Kidder, Pfeiffer, and Teukolsky}}]{Scheel2004}
\bibinfo{author}{\bibfnamefont{M.~A.} \bibnamefont{Scheel}},
  \bibinfo{author}{\bibfnamefont{A.~L.} \bibnamefont{Erickcek}},
  \bibinfo{author}{\bibfnamefont{L.~M.} \bibnamefont{Burko}},
  \bibinfo{author}{\bibfnamefont{L.~E.} \bibnamefont{Kidder}},
  \bibinfo{author}{\bibfnamefont{H.~P.} \bibnamefont{Pfeiffer}},
  \bibnamefont{and} \bibinfo{author}{\bibfnamefont{S.~A.}
  \bibnamefont{Teukolsky}}, \bibinfo{journal}{Phys.\ Rev.\ D}
  \textbf{\bibinfo{volume}{69}}, \bibinfo{pages}{104006}
  (\bibinfo{year}{2004}), \eprint{gr-qc/0305027}.

\bibitem[{\citenamefont{Gleiser et~al.}(2007)\citenamefont{Gleiser, Price, and
  Pullin}}]{Gleiser2007}
\bibinfo{author}{\bibfnamefont{R.~J.} \bibnamefont{Gleiser}},
  \bibinfo{author}{\bibfnamefont{R.~P.} \bibnamefont{Price}}, \bibnamefont{and}
  \bibinfo{author}{\bibfnamefont{J.}~\bibnamefont{Pullin}}
  (\bibinfo{year}{2007}), \eprint{gr-qc/0710.4183}.

\bibitem[{\citenamefont{Hesthaven}(1997)}]{Hesthaven1997}
\bibinfo{author}{\bibfnamefont{J.~S.} \bibnamefont{Hesthaven}},
  \bibinfo{journal}{SIAM J. Sci. Comput.} \textbf{\bibinfo{volume}{18}},
  \bibinfo{pages}{658} (\bibinfo{year}{1997}).

\bibitem[{\citenamefont{Hesthaven}(1999)}]{Hesthaven1999}
\bibinfo{author}{\bibfnamefont{J.~S.} \bibnamefont{Hesthaven}},
  \bibinfo{journal}{SIAM J. Sci. Comput.} \textbf{\bibinfo{volume}{20}},
  \bibinfo{pages}{62} (\bibinfo{year}{1999}).

\bibitem[{\citenamefont{Burko and Khanna}(2007)}]{Burko2007}
\bibinfo{author}{\bibfnamefont{L.~M.} \bibnamefont{Burko}} \bibnamefont{and}
  \bibinfo{author}{\bibfnamefont{G.}~\bibnamefont{Khanna}}
  (\bibinfo{year}{2007}), \eprint{arXiv:0711.0960[gr-qc]}.

\end{thebibliography}

\end{document}